# Apollo 11 and Fundamental Science

A. De Rújula [a, b]

[a] *Instituto de Física Teórica (UAM/CSIC), Autónoma Univ. of Madrid, Spain*
[b] *Theory Division, CERN, CH 1211, Geneva 23, Switzerland*

Half a century after a man first set foot on the Moon it is interesting to revisit the occasion with a measure of hindsight. From the viewpoint of basic science the greatest achievement concerned the implementation of the Nordtvedt test, a precise check of Einstein's strong equivalence principle. A particle physicist may bravely interpret the result as an exquisite measurement of the triple-graviton coupling. Other not so profound experiments were also (unofficially) made in Apollo flights, such as a long-distance test of ESP (yes! extra sensory perception). From a sociopolitical point of view the lesson concerns the feats that can be achieved by a determined and united country… or more than one.

## I. LUNAR LASER RANGING

An estimated one-fifth of the Earth's population –some 3.6 billion people in 1969– watched on live TV the Apollo 11 astronauts as they oddly walked on the Moon. An often-held view is that the only interesting spin-off of the entire enterprise was the development of Moon Boots. Diametrically opposed to that cynical opinion is the fact that Neil Armstrong and Buzz Aldrin, in the two and a half hours that they spent out of the Moon-lander, placed the first passive lunar-ranging laser retro-reflector on our satellite, see Figure 1.

In November 1970, soon after Apollo 11, a French retro-reflector made it to the Moon onboard the Soviet probe Lunokhod 1[1]. This devise was so endearingly Star Wars-like that the temptation to show a model of it is irresistible, thus Figure 2. Three other reflectors were placed on the Moon by Apollo 14 and 15 and Lunokhod 2.

---

[1] In Russian Lunokhod means lunar walk.

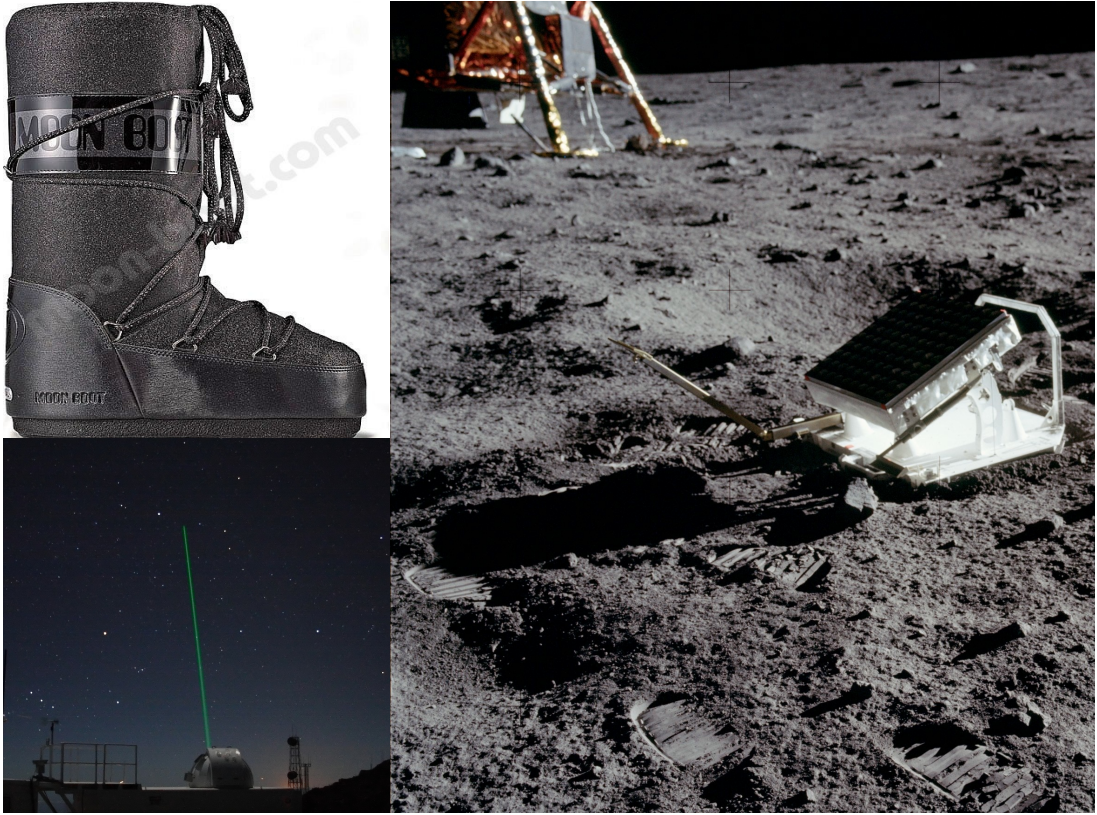

Figure 1: A Moon Boot, the retro-reflector and a laser pointing to it.

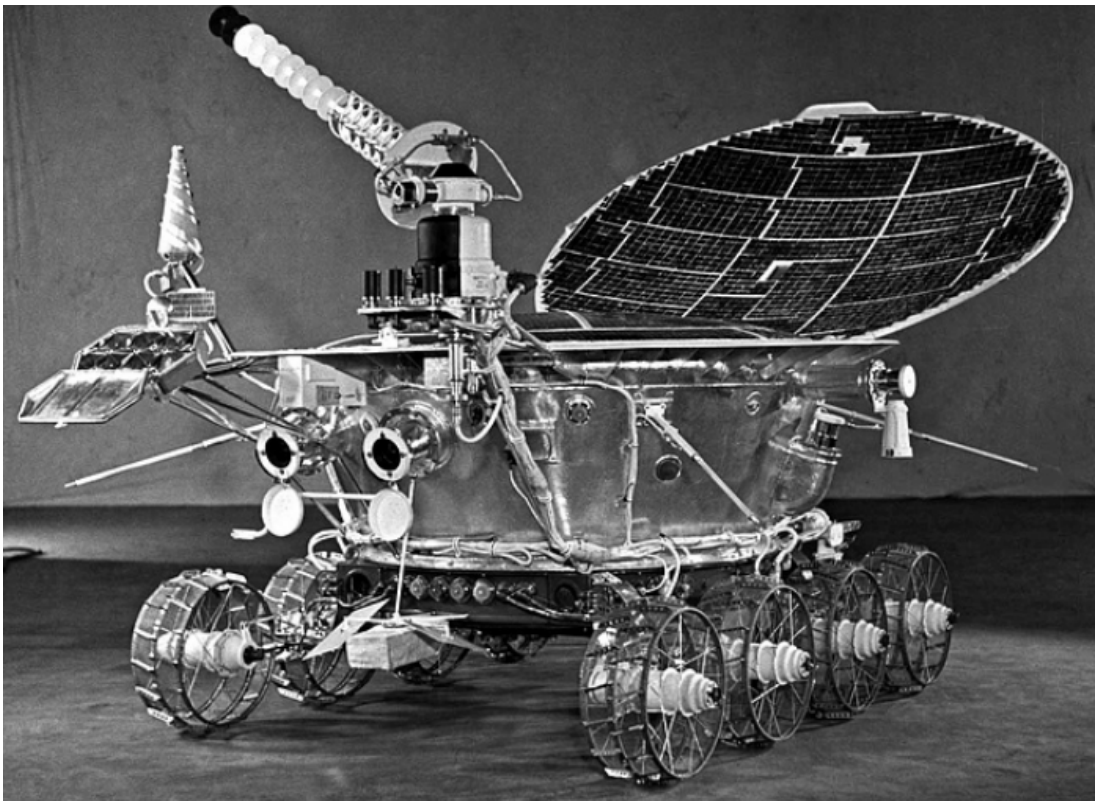

Figure 2: Lunokhod 1. The French retro-reflector is in the open box on the upper left.

The above-cited retro-reflectors were part of the instruments used to make a precise and beautiful test of the strong equivalence principle of General Relativity. These sophisticated mirrors are not tiny gadgets, the one in Figure 1 measured $59 \times 67 \times 27.5$ cm$^3$. It had a mass of a few pounds, but a considerable volume, not an entirely negligible fraction of the space available in the Apollo Lunar Module. The fact that NASA, on the pioneering Apollo 11 mission, knowingly "sacrificed" a fraction of its weight- and space-lifting might to what turned out to be fundamental physics [1] deserves an extremely loud **hurrah**.

The plan to place a retro-reflector on the Moon was initiated by James E. Faller in a note [2] to Robert H. Dicke, the beginning of which is shown in Figure 3. This was the application to the Moon of a similar program to measure distances to artificial satellites, originally proposed in a preprint of July 15$^{th}$, 1959 by W.H. Hoffman, R. Krotkov and Dicke [3].

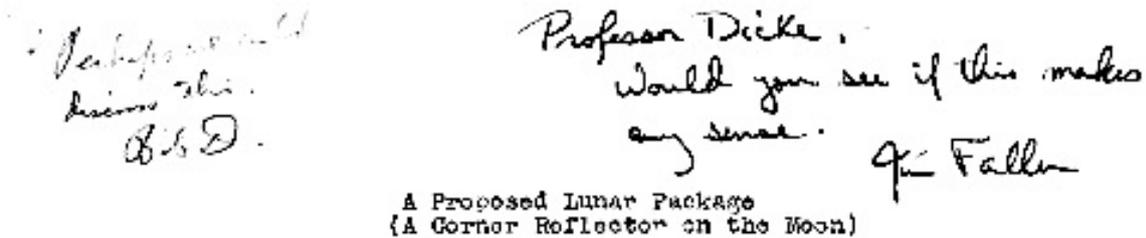

Figure 3. Frazer's note to Dicke: *"Would you see if this makes any sense"*. Notice also that the original plan was for an unmanned mission.

In a first approximation lunar ranging (the measurement of the Earth-Moon distance as a function of time) is quite simple. A strong ultra-short laser pulse is directed to the mirror[2], to which it arrives with an aperture of circa 6.5 km. The mirror is "retro" in the sense that the light bounces back in the direction from which it came, rather than the complementary one (how to build such a mirror is left as an easy exercise for the reader). About one out of $10^{17}$ of these monochromatic photons gets back and may be detected. Half the time spent on the round trip, times the velocity of light, *c*, is the quasi-instantaneous distance from the laser-emitter to the reflector. After several decades of sporadic measurements the current uncertainty of the distance determinations is below 1 cm. A first "trivial" result is that the Moon recedes from the Earth about 3.8 cm per year, due to tidal friction.

## II. *SCIENCE:* THE NORDVEDT TEST

In 1968 Kenneth Nordtvedt, following an earlier remark by Robert Dicke, analysed in detail how to extend Galileo's supposed experiment at the leaning tower of Pisa to check whether or not the Earth and the Moon "fall" towards the Sun with the same acceleration [1]. Nordtvedt added a crucial detail, which we shall recall anon, regarding the gravitational self-masses of the two lighter bodies.

If our planet and its satellite do not "suffer" the same Sun-wards acceleration, the Moon's orbit would be "polarized" in the direction of the Sun[3]. When the Earth, Moon and Sun are aligned the effect is maximal. For any other "synodic" angle (between the Moon's and Sun's directions as seen from the Earth) the effect varies in a sinusoidal manner.

It is easy to guess that the hypothetical radial modification of the Moon's orbit would be, to a first approximation, proportional to the difference of accelerations times the cosine of (2πt/T), with T the synodic period (of the successive alignments of the three bodies[4]). Figuring out the coefficient of proportionality is less trivial [1]. Having a specific period or frequency is a very welcome feature in the search for a hypothetical effect: a good fraction of high-precision physics would cease to exist in the absence of the teachings of Jean-Baptiste Joseph Fourier. In this case how to transform information from positions in space to frequencies.

---

[2] The most advanced lunar laser ranging station (APOLLO in New Mexico) uses a 3.5-meter telescope and a 532 nm Nd:YAG laser. Its pulses have an energy of 115 mJ and a duration of 0.1 ns (~ 3 cm of light travel).

[3] The word used in the trade, "polarization" may be a bit obscure. It means a stretching of the Moon's orbit in the direction of the Sun. Trademarked expressions induce misinterpretations. Gauge, for instance, means a measuring device. Particle physicists are fond of "gauge symmetries", which are not symmetries and refer to something that cannot be measured.

[4] In case a reminder is not unwelcome. A sidereal year, $Y = 365.25636$ days, is the time between two consecutive equal positions of the Earth in its orbit around the Sun, as seen by ET from a distant star. A sidereal lunar period, $\tau = 27.32158$ days, is the time for the Moon to complete a revolution around the Earth, also as seen by ET. A little drawing and thought result in $T = 1/(\tau^{-1} - Y^{-1}) = 29.53049$ days, the synodic interval between two consecutive new Moons.

### IIa. Testing the weak equivalence principle

The mass of an atom, say Hydrogen, is the sum of the masses of its electron, its proton and the negative contribution due to the electromagnetic binding between the two charged constituents. A (negative) binding energy ΔE corresponds to a contribution $\Delta E/c^2$ to the mass. The cited "mass" could be inertial (the *m* in $E = mv^2/2$ with *E* the non-relativistic kinetic energy) or gravitational (the *m* in $F = GmM/R^2$ with *G* Newton's constant and *F* the weight of the object here, if *M* and *R* are the mass and radius of the Earth). The stated identity between *m&m* is an assumption: the **weak equivalence principle**. It is tested to exquisite precision by comparing the two types of masses for materials of different substances, all of which have the same constituents (electrons, protons and neutrons) but different atomic, molecular and/or solid-state binding energies [4].

Two objects with different ratios of their gravitational to inertial mass would fall with a difference, Δg, in their accelerations. Some recent limits from laboratory [5] or satellite [6] experiments, normalized to the (almost) common g, are:

$$\Delta g[\text{Be}, \text{Ti}]/g < (0.3 \pm 1.8)\, 10^{-13} \qquad [5]$$

$$\Delta g[\text{Be}, \text{Al}]/g < (-0.7 \pm 1.3)\, 10^{-13} \qquad [5]$$

$$\Delta g[\text{Ti}, \text{Pt}]/g < (-1 \pm 9[\text{stat}] \pm 9[\text{syst}])\, 10^{-15} \qquad [6]$$

These limits to departures from the weak equivalence principle are extremely impressive.

### IIb. Testing the strong equivalence principle

Let *r* be the radial distance of a volume element of an object from its centre and ρ its local density. The energy of the gravitational attraction between any two volume elements is[5]:

$$\Delta E_G = -\int \frac{G\,\rho(\vec{r})\,\rho(\vec{r}')}{2\,|\vec{r}-\vec{r}'|}\,d^3r\,d^3r'$$

---

[5] For a uniform-density sphere the gravitational self-energy is $3\,G\,M^2/(5\,R)$.

and, in the sense to be made explicit, it *ought to* contribute to the rest mass of the object an amount $\Delta M = \Delta E_G/c^2$. Define an object's mass, without this contribution, as $M_0$. The **strong equivalence principle** is the statement that the pull of gravity acts on $M = M_0 + \Delta M$ (it adds the "weight" of gravity to the weak equivalence principle, thus stating that gravitational self-energies also gravitate). The strong principle is respected by general relativity, while alternative theories of gravity allow for $M = M_0 + \eta \Delta M$, with $\eta \neq 1$ [4].

In an atom the gravitational binding energy is totally negligible. But gravitational forces always have the same sign (except for the cosmological constant, if naively interpreted as a force) and they increase linearly with mass. The Earth and the Moon are large enough for their gravitational self-masses not to be desperately negligible. Their ratios to the objects' total mass are approximately $-4.64 \times 10^{-10}$ and $-2.0 \times 10^{-11}$ for the Earth and the Moon, respectively, see Figure 4.

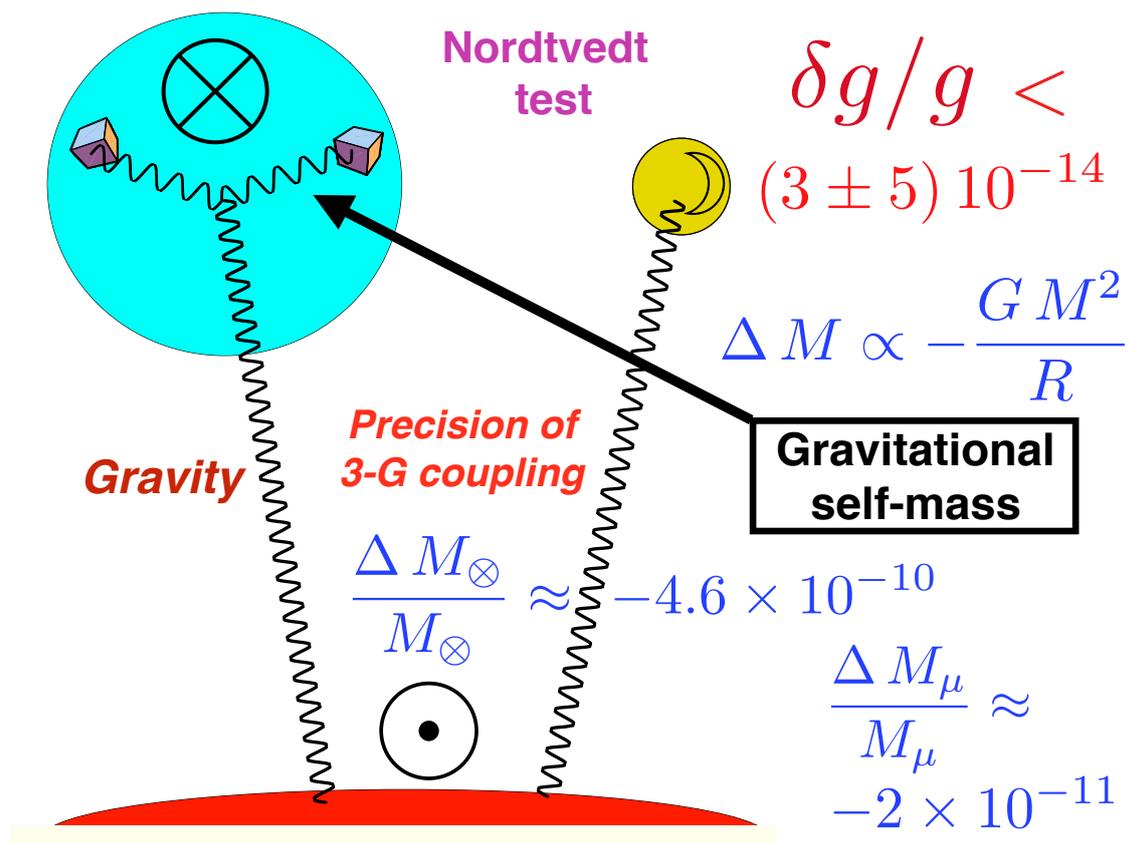

Figure 4: The Moon-Earth-Sun Nordtvedt test of the strong equivalence principle.

The lunar-ranging observations constrain the equality of the accelerations with which the Moon and the Earth fall towards the Sun to the level:

$$\Delta g[E, M]/g < (3 \pm 5) \, 10^{-14}$$

Given the estimated gravitational self-masses of the Earth and Moon, the non-observation of a synodic "Nordtvedt polarization" of the Moon's orbit results in a value, or the corresponding upper limit, of the strong-equivalence-violating parameter $\Delta \eta \equiv \eta - 1$ [4]:

$$\Delta \eta = (-0.2 \pm 1.1) \, 10^{-4}$$

This is the level to which we know from lunar ranging that, as in Einstein's theory, "gravity gravitates". More in detail, the gravitational self-masses of the Earth and the Moon —represented in Figure 4 by the wiggly lines between two volume elements— are pulled by the Sun's gravity just as much as the bulk of their masses are.

Even if individual gravitons have not been detected —and will not be for the foreseeable future— a particle theorist [7] would be tempted to interpret the wavy-lines' vertex in Figure 4 as the *triple-graviton coupling*[6]. A fundamental difference between electrodynamics and gravity is that the second is "nonlinear", in the sense that the carriers of the force, unlike photons, also act as sources of gravity. Yet another way to state that gravity gravitates is that general relativity is a *non-abelian gauge theory*. The same statement applies to the electroweak Standard Model of quarks and leptons.

The case of Quantum Chromodynamics (QCD, the theory that describes the strong interactions of quarks and gluons) is entirely akin to General Relativity, in the sense that gluons have "color charges" and couple to gluons. The irony is that, in spite of the fact that at accessible energies gravity is the weakest and QCD the strongest of the known forces between elementary particles, the triple-graviton coupling is measured to be "what it should be" to a three orders of magnitude better precision than the triple gluon coupling. Testing QCD to an astronomically equivalent level of accuracy would be no mean feat.

### IIc. Further tests

There is a very unlikely-sounding loophole in the Nordtvedt test as we discussed it. Conspiracy theories being so much in vogue, it could be said that the different chemical mass composition of the Earth and the

---

[6] This is for the same reason that the Fourier transform of the one-photon exchange term (in the Hamiltonian) between electrical charges is their Coulomb interaction energy.

Moon conspire, via unknown feeble composition-dependent forces, to invalidate the conclusion that the strong equivalence principle is tested to the stated precision. To distinguish weak from strong equivalence-principle effects the Eöt-Wash group at the University of Washington performed a torsion balance experiment using test masses of similar composition to the Earth and Moon [8]. Combining the torsion balance results with the latest lunar ranging analyses led to the result [9]:

$$\Delta\eta = (4.4 \pm 4.5)\, 10^{-4},$$

which is only slightly less restrictive than the non-conspiratorial one.

The effects we have discussed are not the only ones required to extract a test of general relativity. The analysis of the data requires, among many others, corrections due to geophysical and rotational effects for the Earth and the Moon, in addition to orbital effects. Lunar ranging can also be used to extract other tests, such as a hypothetical variation with time of Newton's constant [10]. Recent results are:

$$\dot{G}/G = (6 \pm 7) \times 10^{-13}/\text{year} \quad [11],$$
$$\dot{G}/G = (2 \pm 7)\, 10^{-13}/\text{year} \quad [12],$$
$$\ddot{G}/G = (4 \pm 5)\, 10^{-15}/\text{year}^2 \quad [12].$$

These limits imply a < 1% variation of $G$ over the age of the universe [12].

The Apollo 11 astronauts also planted on the Moon a seismometer. Though not as "fundamental" as the science based on the retro-reflector, what was learned from this device and similar ones is quite interesting. Moon-quakes, mainly generated by crashing meteorites, allowed to study the insides of the Moon and to find, for instance, that, not unlike the Earth, it has an iron core. Even Aldrin's iconic Apollo 11 boot-print photo had a scientific interest: it revealed much about the local lunar soil, including its fine-grained nature, its cohesiveness, and its ability to pack tightly together.

### III. LEGAL ISSUES AND UNSCIENTIFIC NONSENSE

There were other aspects of the Apollo program more controversial than lunar ranging. Frank Borman, Jim Lovell and Bill Anders, the astronauts in Apollo 8, circled the Moon and on Christmas Eve 1968 they took the unforgettable Earthrise picture (selfies were not yet pervasive). They also read a passage from the Old Testament. Madalyn Murray O'Hair --a professional atheist of sorts— sued NASA, asserting that her First Amendment rights had been violated. The judge dismissed the suit and the Supreme Court declined to hear it due to lack of jurisdiction [13].

Raising the American flag on the Moon, as the Apollo 11 crew did, also turned out to be controversial. The concern was that planting this particular flag on the Moon may make it look like the Americans were taking control of our satellite, which would have been a violation of the Outer Space Treaty[7]. Eventually, however, it was decided to "rise" a stiff American flag, while leaving a plaque to emphasize that the astronauts "came in peace for all mankind."

The extent to which the flag debate involved more lawyers than astronauts is clear from the subsequent developments. In 1969 NASA's appropriation bill stated *"the flag of the United States, and no other flag, shall be implanted or otherwise placed on the surface of the Moon, or on the surface of any planet, by the members of the crew of any spacecraft making a lunar or planetary landing as a part of a mission under the Apollo program or as a part of a mission under any subsequent program, the funds for which are provided entirely by the Government of the United States."* In an attempt to respect the Outer Space Treaty, the bill made sure to note that *"This act is intended as a symbolic gesture of national pride in achievement and is not to be construed as a declaration of national appropriation by claim of sovereignty."*

Experiments perhaps less reputable than lunar ranging were also performed in Apollo missions [14]. Believe it or not, I am referring to ESP. Yes, Extra Sensory Perception, indeed! During his voyage to the Moon on Apollo 14, Edgar Mitchell secretly performed an ESP experiment. While his fellow crewmembers Alan Shepard and Stuart Roosa slept, Mitchell took out a collection of cards and spent a few minutes concentrating on a random series of "Zener" symbols. Back on Earth, a team of psychics tried to read his thoughts and write down the order of the sequence. The group reportedly guessed the right numbers 51 times out of 200, which Mitchell described as "results far exceeding anything expected" [15]. Hard to believe!

Actually, there was a prearranged time when Mitchell and his friends would do their tests, but problems prevented things from going as planned. As a result, the recorded guesses on Earth were made *before* Mitchell went through the trials. No problem. In his published paper on the experiment, Mitchell just changed its goal to a study of *precognition* [16].

---

[7] The Outer Space Treaty**,** formally *Treaty on Principles Governing the Activities of States in the Exploration and Use of Outer Space, Including the Moon and Other Celestial Bodies*, (1967), was meant to bind the parties to use outer space only for peaceful purposes. It came into force on Oct. 10, 1967, after being ratified by the United States, the Soviet Union, the United Kingdom, and several other countries.

# IV. SPINOFFS OR THE APOLLO PROGRAM

To continue with a less grim note, let me attempt to undo the initial allegation that the only technological spinoff of the Apollo program was the Moon boot. In a NASA webpage dedicated to this subject [17], one finds, amongst a few others, the following items:

∗ Cooling suits and flame-resistant materials, now used by racing drivers, nuclear reactor technicians and others.

∗ Chemical processes for toxic waste removal in recycled fluids now used in kidney dialysis.

∗ Apollo water purification technology currently used in some community water supply systems.

∗ A space cardiovascular conditioner that evolved into physical therapy and athletic development machines.

* The insulating barriers made of metalized foil laid over a core of propylene or mylar, which protected astronauts and their spacecrafts' delicate instruments from radiation and heat, are now found in common home insulation.

* Freeze-dried foods that preserve nutrients [not the most appetizing spinoff].

* The same fabric used in Apollo-era space suits has evolved into an environmentally friendly building material, a Teflon-coated fiberglass.

* Retro-reflectors similar to the lunar-ranging ones are used to detect hazardous gases in some of the many places that generate them.

* To avoid the outgassing of lightweight metals to result in the malfunctioning of lubricants, dry lubricant coatings were developed. They are now employed in laser manufacturing and… pizza making.

Estimates of the total cost of the Apollo program vary by up to a factor of two. Two presumably reliable sources, Forbes [18] and the BBC [19], place it at 152 and 175 billion in today's dollars, respectively. Given these numbers, the technical spinoffs of the program appear to be rather meager, even considering the one cited last in the above list. The value of a test of the strong equivalence principle cannot be stated in dollars, but others areas of research, such as condensed matter and particle physics, have no doubt had a very much larger spinoff to cost ratio. But nobody, not even graphene or Higgs bosons can compete with NASA on publicity and impact on the media, with the exception of one medium of the media: the World Wide Web, a spinoff from CERN.

## V. WINDUP

Speaking to Congress and the Nation, President Kennedy said on May 25, 1961: *"I believe that this nation should commit itself to achieving the goal, before this decade is out, of landing a man on the Moon and returning him safely to the Earth. No single space project… will be more exciting, or more impressive to mankind, or more important… and none will be so difficult or expensive to accomplish."* [20].

Are there any current goals equally *"exciting, impressive and important"*? The two obvious answers are even more difficult and expensive than landing a man on the Moon, and are certainly more urgent.